\documentclass[sigconf, authorversion=true]{acmart}

\usepackage[utf8]{inputenc}
\usepackage{url}
\usepackage{caption}
\usepackage{breqn}
\usepackage{amsmath}
\usepackage{bm}
\usepackage{xcolor}
\usepackage{graphicx}
\usepackage{subcaption}

\newcommand{\ostyle}{\oldstylenums{100}\textsc{style}}

\AtBeginDocument{%
  \providecommand\BibTeX{{%
    \normalfont B\kern-0.5em{\scshape i\kern-0.25em b}\kern-0.8em\TeX}}}

\begin{document}

\title[Feature-Wise Transformations and Local Motion Phases]{Real-Time Style Modelling of Human Locomotion via Feature-Wise Transformations and Local Motion Phases}

\author{Ian Mason}
\affiliation{%
  \institution{University of Edinburgh}
  \country{United Kingdom}
}
\email{ianxmason@gmail.com}

\author{Sebastian Starke}
\affiliation{%
  \institution{University of Edinburgh}
  \institution{Electronic Arts}
  \country{United Kingdom}
}
\email{sebastian.starke@ed.ac.uk}

\author{Taku Komura}
\affiliation{%
  \institution{University of Edinburgh}
  \institution{Hong Kong University}
  \country{Hong Kong}
}
\email{tkomura@ed.ac.uk}

\begin{abstract}
  Controlling the manner in which a character moves in a real-time animation system is a challenging task with useful applications. Existing style transfer systems require access to a reference content motion clip, however, in real-time systems the future motion content is unknown and liable to change with user input. In this work we present a style modelling system that uses an animation synthesis network to model motion content based on local motion phases. An additional style modulation network uses feature-wise transformations to modulate style in real-time. To evaluate our method, we create and release a new style modelling dataset, \ostyle,  containing over 4 million frames of stylised locomotion data in 100 different styles that present a number of challenges for existing systems. To model these styles, we extend the local phase calculation with a contact-free formulation. In comparison to other methods for real-time style modelling, we show our system is more robust and efficient in its style representation while improving motion quality.
\end{abstract}



\begin{teaserfigure}
  \includegraphics[width=\textwidth]{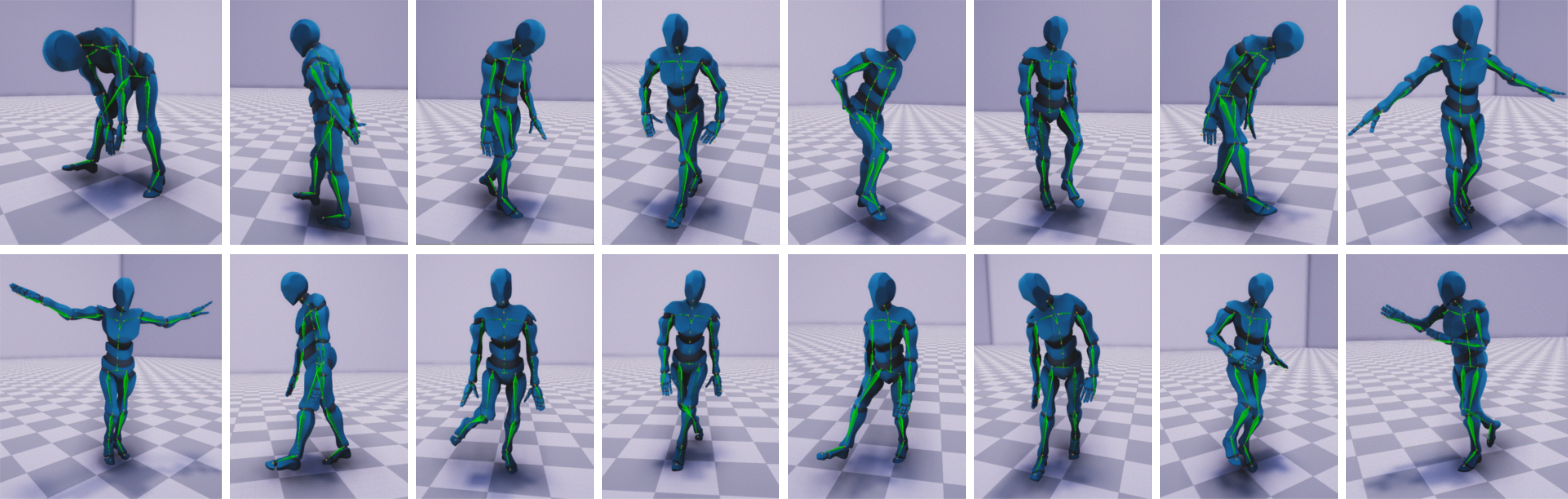}
  \caption{Modelling large numbers of locomotion styles.}
  \Description{Modelling large numbers of locomotion styles.}
  \label{fig:teaser}
\end{teaserfigure}

\maketitle

\section{Introduction}

Modelling the different ways in which people can move has many applications in computer graphics such as changing a video game character's gait after being injured, or conveying character personality through different walking styles in animated movies. Recently, several works for data driven \emph{style transfer} have achieved high quality results by learning to extract style information from motion capture data and using this to alter the movements of humanoid characters \cite{aberman2020unpaired, holden2017style, smith2019efficient, xia2015style, yumer2016spectral}. Such works provide \emph{offline style transfer} that explicitly maps the style of a given motion clip onto the content of a different clip (walking/running, direction, trajectory etc.). These methods often make use of explicit timing adjustments or foot contact matching between content and output stylised clips and cannot easily be applied for \emph{real-time style modelling} in systems where the future timing and contacts are unknown and may change rapidly as user input changes.

Modern data driven \emph{animation synthesis} systems \cite{holden2017pfnn, ling2020motionvae, starke2019nsm, starke2020local, zhang2018mann} do not need to make such timing adjustments, instead using neural networks to model animation frames autoregressively. These methods automatically capture how timing, stepping and bone positions change conditioned on past frames \emph{and} user input. However, most of these systems do not allow for multiple modalities of movement and require retraining to model new styles or new characters. Those works that do model style explicitly \cite{holden2017pfnn} make use of methods which cannot easily extend to new styles and do not provide optimal quality (e.g.\ one-hot labelling, see Section~\ref{sec:evaluation}). 

\emph{Style modelling} sits between style transfer and animation synthesis. The aim of style modelling is to model motion conditioned on user input, as in animation synthesis, but to additionally allow for the changing of style on the fly \cite{holden2017pfnn, holden2020learned, mason2018style}. Unlike style transfer, style modelling systems work online with arbitrary lengths of content determined by the user input. Unlike most animation synthesis systems, style modelling systems are explicitly designed for modelling large numbers of motion styles and require an element of animation editability not currently present in these fully-neural systems. A good style modelling system should be able to: well model stylised locomotion in real-time conditioned on user input; model an arbitrary number of styles; allow for new styles to be added easily (via generalisation or low-cost fine-tuning), and interpolate/transition well between styles. 

In this paper, we propose a novel, real-time approach for style modelling.  We do not require paired or aligned data, instead motions are stylised on a per-frame basis by conditioning on learned style representations. We design a framework consisting of an \emph{animation synthesis network}, based on a local phase network (LPN \cite{starke2020local}), augmented with a \emph{style modulation network} which uses feature-wise linear modulation (FiLM \cite{dumoulin2018feature-wise, perez2018film}) to transform the hidden layers of the animation synthesis network. This framework allows us to learn compact representations for many different styles while maintaining the interactive control of a real-time system. In order to utilise the LPN with stylised locomotion, which does not contain the contact information required to create local phases \citep{starke2020local}, we additionally design a contact-free local phase function.

To evaluate the strengths and limitations of our approach, we create and release the \ostyle\ dataset containing 100 different performative styles of locomotion and over 4 million frames of motion capture data. Combining this dataset with our style modelling framework allows us to model a wider variety of styles than previously shown in the literature, to synthesize different modes of locomotion (walking forwards and backwards, sidestepping etc.), and to create continuous transitions between styles. The \ostyle\ dataset also helps us to analyse areas of limitation for existing works and to improve over systems that have been limited by available $3D$ stylised data.  

The main contributions provided by this work are to:
\begin{itemize}
    \item Create an improved style modelling system, using a novel combination of animation synthesis and style modulation.
    \item Release the \ostyle\ dataset with $\sim$4 million mocap frames.
    \item Extend the local phase formulation of \cite{starke2020local} to the contact-free setting, increasing the applicability of the LPN.
    \item Demonstrate modelling of a larger number of styles than previously shown by using a low cost style representation.
\end{itemize}

\section{Related Work}
Following the introduction, we review related work first on animation synthesis and then on style transfer and modelling. 

\vspace{-2mm}
\subsection{Data-Driven Motion Modelling}
Many classic machine learning techniques have been applied for modelling human motion from data including, blending radial basis functions \cite{rose1998verbs}, Gaussian processes \cite{ikemoto2009edits, mukai2005geostatistical} and PCA \cite{safonova2004lowdim}. More recently, autoregressive neural network methods have been developed to predict future motion frames from given input frames \cite{fragkiadaki2015, li2017aclstm, Martinez_2017_CVPR, butepage2017}. Some authors have also used reinforcement learning to model motion in physics-based systems \cite{peng2017deeploco, yu2018lowenergy, peng2018mimic, fussell2021supertrack}, although animation quality still tails recent kinematic approaches.

More relevant to this work is locomotion modelling for controllable animation synthesis \cite{holden2016framework, ling2020motionvae, henter2020moglow}. One successful method is the Phase-Functioned Neural Network \cite{holden2017pfnn} which alters neural network weights dependent on a global motion phase. This creates implicit data alignment which allows for the generation of high quality animation. 
This has also been extended to blending sets of expert weights to model quadruped motion and character interactions with a scene \citep{zhang2018mann, starke2019nsm}. Recently, \citet{starke2020local} have proposed local motion phases to better model asynchronous limb movements; our method builds on top of this system as many motion styles lack a coherent global phase. While successful for specific tasks, these methods do not learn general motion features so lack the ability to extend to unseen movements without retraining the whole system.

\vspace{-2mm}
\subsection{Style Transfer and Modelling}

Stylisation techniques have a long history in computer vision and graphics with applications to different mediums. For example, since the development of a neural approach by \citet{gatys2015style}, many methods have been designed that apply transformations to hidden units of convolutional networks to stylise images \cite{dumoulin2017style, huang2017adain, huang2018multimodal}. Style transfer is also closely related to the wider task of domain adaptation \citep{Storkey09, rebuffi2018, atapour2018real, hou2020source, eastwood2021sourcefree, eastwood2021unitlevel}.

For animation data, whilst some methods have used hand designed approaches to create a correspondence between content and style \cite{hsu2005, yumer2016spectral}, most modern methods have focused on learning from data \cite{taylor2009rbm, xia2015style, holden2017style, du2019style, smith2019efficient, dong2020adult2child, Wen_2021_CVPR, park2021diverse}. These style transfer methods however, require the motion content to be specified upfront, usually via a clip of motion capture data, and thus are unsuitable for interactive applications with high level user control.

For controllable systems, \citet{holden2017pfnn} use a one-hot style representation to learn a small number of motion styles. However, this requires the total number of styles to be known before training and, with large number of styles, many of the parameters are used simply for multiplication with the label. Learned motion matching \cite{holden2020learned} is able to efficiently model stylised motion but learns no explicit style representation, rather using a general feature matching approach that increases in cost with the dataset size. Most similar to our work, \citet{mason2018style} showed that style could be modelled in real-time interactive systems by altering the hidden features of neural network layers using residual adapters \cite{rebuffi2017residual}. \looseness-1 They did not however, try to learn a general style parameterisation function. \citet{aberman2020unpaired} use a similar approach, transforming hidden units to transfer style onto given content clips from $2D$ or $3D$ data, however, they did not experiment with real-time controllable systems.   

\begin{figure*}[t]
\centering
\includegraphics[trim={0 1.5cm 0 0.5cm}, clip, width=\linewidth]{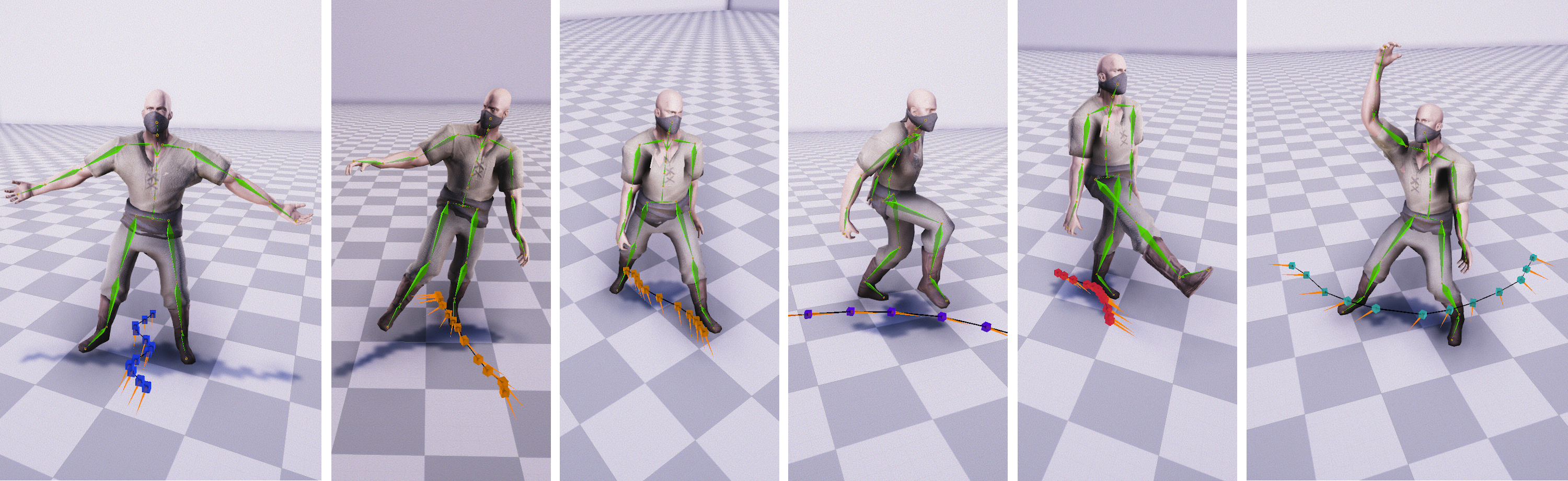}
\caption{Six of the styles from the \ostyle\ dataset. From left to right: \emph{Star} - Backwards Walk; \emph{Drunk} - Forwards Walk; \emph{Drag Right Leg} - Forwards Walk; \emph{Roadrunner} - Forwards Run; \emph{Kick} - Backwards Walk, and \emph{Wild Arms} - Sidestep Run. The \emph{Drunk} style is stochastic, \emph{Drag Right Leg} is asymmetric and \emph{Wild Arms} has joints moving with different phase cycles.}
\label{fig:data}
\vspace{-2mm}
\end{figure*}

\section{Stylised Locomotion Data}
\label{sec:data}
We now describe the data created for this work. We begin by presenting our new, large dataset of 100 locomotion styles and then discuss the specifics of processing this dataset for use in our models. This dataset is released alongside this paper.

\vspace{-2mm}
\subsection{The 100STYLE Dataset}
Much of the recent work in data-driven style transfer has used relatively small datasets. The lack of availability of large datasets has led to the development of augmentation techniques \cite{lee2018interactive}, methods that work with small amounts of data \cite{mason2018style}, methods that try and leverage 2D data \cite{aberman2020unpaired} and domain specific solutions \cite{yumer2016spectral}. Those methods that are purely data-driven are often only able to evaluate on a small number of globally cyclic styles \cite{holden2017style,smith2019efficient}. Our dataset is designed for modern data hungry techniques and contains 100 styles of locomotion each with a large number of frames of motion capture data. Whilst there exists datasets which contain a large number of frames for each style they tend to have a limited number of styles \cite{ionescu2014h36m,xia2015style}. Datasets with a large variety of motions tend to have relatively few frames of data per style \cite{Gross01thecmu}. Our dataset can combine the benefits of both approaches.

The locomotion styles in our dataset are designed to cover a range of movements that may prove challenging to model with a single system. Whilst some styles are variations on a neutral walk we also capture styles that are stochastic, asymmetric, and do not have a clear global phase. We leave discussion of what constitutes a style to Section~\ref{sec:discussion}, Figure~\ref{fig:data} shows some styles captured in \ostyle.

The data is captured at $60$fps in a 4.5m x 4.5m capture area using a single actor of height 182cm and Xsens motion capture technology. The skeleton contains 28 joints, including full-body position and rotation data for each joint, but excluding finger transformations. Each style covers idling, walking and running movements and is captured according to a pre-prepared script, see Table \ref{tab:breakdown} for the breakdown of the number of frames per gait type. Full information including a list of style descriptions, frames per style, and the script for motion capture can be found in the supplementary material. 

\begin{table}[]
\caption{A summary of the quantity of different gaits captured in the \ostyle\ dataset}
\vspace{-2mm}
\begin{tabular}{llll}
\toprule
Motion Gait      & Frames    & Time(m)  & Ratio(\%) \\ 
\midrule
Backwards Run    & 434,329   & 120      & 10.7       \\
Backwards Walk   & 768,574   & 213      & 19.0       \\
Forwards Run     & 414,911   & 115      & 10.2       \\
Forwards Walk    & 777,640   & 216      & 19.2       \\
Sidestep Run     & 250,980   & 70       & 6.2        \\
Sidestep Walk    & 541,041   & 150      & 13.3       \\
Idling           & 81,685    & 23       & 2.0        \\
Transitions      & 786,818   & 218      & 19.4       \\
\midrule
Total            & 4,055,978 & 1125     & 100.0      \\
\bottomrule
\end{tabular}
\label{tab:breakdown}
\vspace{-4mm}
\end{table}

\vspace{-2mm}
\subsection{Data Processing}
\label{sec:dataproc}
In order to use the \ostyle\ dataset we first remove the T-poses and any non-stylised locomotion from the motion capture before mirroring symmetric styles to increase the data quantity and extracting a number of features. We create three datapoint types: input frames, output frames that we aim to predict, and clips of stylised locomotion from which we aim to learn the style.

The input frame, $\mathbf{x} \in \mathbb{R}^{348}$, consists of: $2 \times 12$, $x$ and $z$ trajectory positions, representing the projected root position of the character every 10 frames for 1 second in the future and past; similarly $2 \times 12$ $x$ and $z$ trajectory facing directions; a $75$ dimensional vector containing $x$, $y$, $z$ positions for each of the $25$ joints in our skeleton (excluding the toes and top of head joint in the original motion data, seen in Figure \ref{fig:teaser}); another $75$ dimensional vector containing $3D$ velocities for each of the joints, and $150$ dimensions for quaternion forwards and upwards vectors for the rotations (as in \cite{zhang2018mann}). All features are extracted in the character's local coordinate frame to provide invariance to global location. We additionally extract an 8-dimensional local phase representation as explained in Section~\ref{sec:anim_syn}.

The output frame, $\mathbf{z} \in \mathbb{R}^{342}$, consists of the, $2 \times 6$, $x$ and $z$ trajectory positions and directions for one second in the future, along with the same $300$ dimensions for joint positions, velocities and rotations as the input frame, with all values taken relative to the character's position in the input frame. In this way we can model the character's global root offset between frames. We also include an approximate foot contact label (further described in Section~\ref{sec:anim_syn}), with the remaining $16$ dimensions for the local phases and local phase updates required to run the LPN autoregressively.

The input clip, $\mathbf{y} \in \mathbb{R}^{240 \times 300}$, is a $240$ frame sequence containing the relative joint positions, velocities and rotations for each frame. We do not include the trajectory or global offset as we don't wish to model the specific (global) motion in the clip.

\begin{figure*}[t]
\centering
\captionsetup{width=.9\linewidth}
\includegraphics[width=.9\linewidth, trim={0 1.3cm 0 1.1cm}, clip]{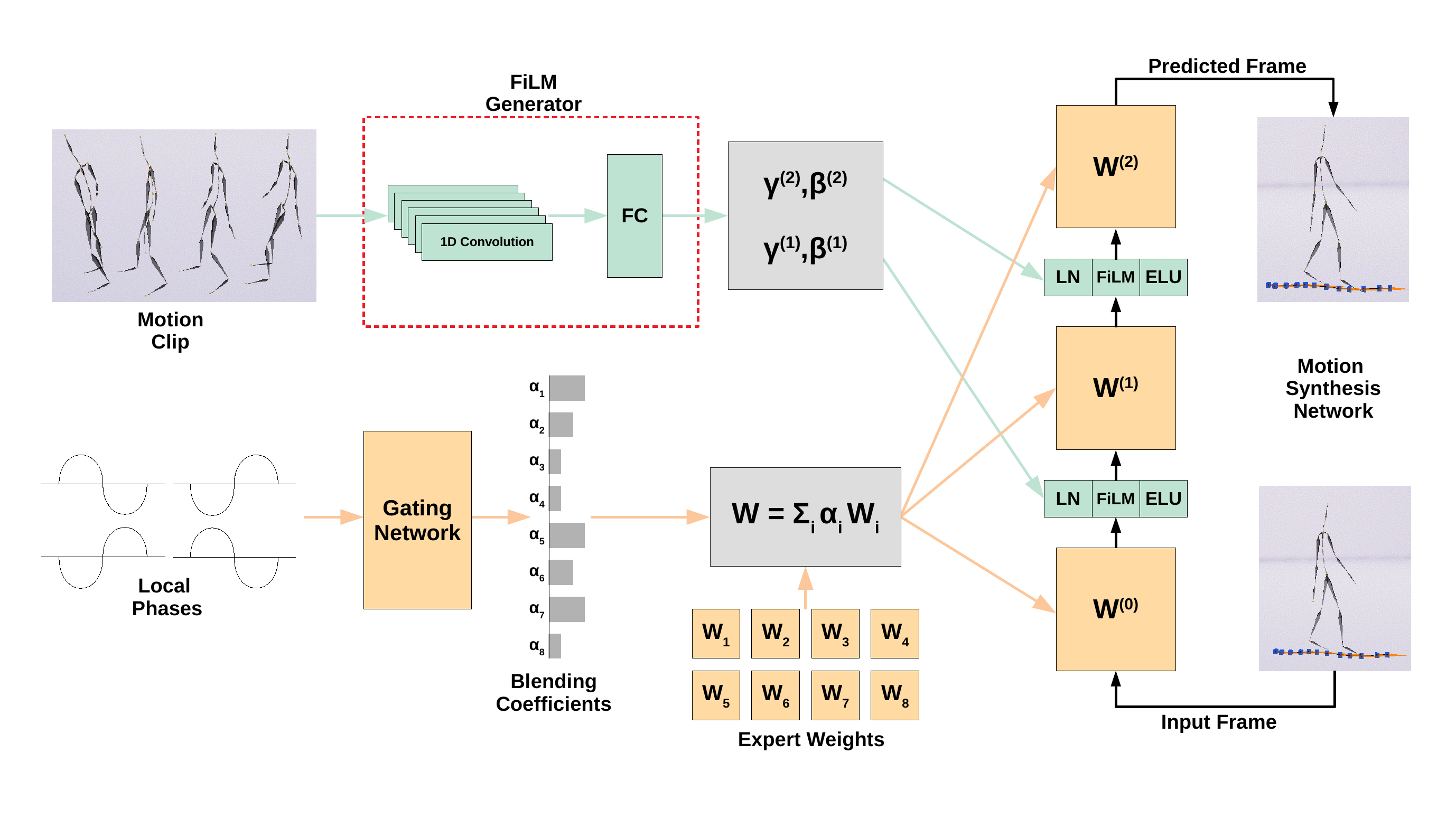}
\caption{System Overview. Animation is generated with an animation synthesis network (peach) that uses local phases along with FiLM parameters from a style modulation network (green). LN is layer normalisation \cite{ba2016layer}.}
\label{fig:architecture}
\vspace{-2mm}
\end{figure*}

\section{Style Modelling Methodology}
\label{sec:method}
In this section we describe a system for style modelling. \looseness-1 Our system contains an animation synthesis network that predicts the next frame of motion from a given input frame, and a style modulation network that extracts style information from clips of motion data and uses this to modulate the hidden layers of the animation synthesis network. A summary of our architecture is shown in Figure~\ref{fig:architecture}.

\vspace{-2mm}
\subsection{Animation Synthesis Network}
\label{sec:anim_syn}

We use a state of the art network based on a gated experts approach designed to capture local movements in locomotion \citep{starke2019nsm, starke2020local, zhang2018mann}. This network consists of two parts, firstly a gating network, $\bm{\Omega}$, that takes as input some local signal and outputs blending coefficients $\{\alpha_i\}_{i=1}^K$. These coefficients are used to linearly blend $K$ sets of expert weights $\{\mathbf{W}_i\}_{i=1}^K$, with each set of expert weights themselves representing one possible configuration of neural network weights. This forms the parameters for the second network, the motion synthesis network $\bm{\Phi}$, as $\mathbf{W} = \sum_i \alpha_i\mathbf{W}_i$. We denote the parameters of the $l^{th}$ layer of the motion synthesis network as $\mathbf{W}^{(l)}$.

\vspace{-1mm}
\subsubsection{A Hybrid Approach to Local Phases}
When defining phase cycles for locomotion it is common to use a global phase defined by the foot contacts (one phase cycle is one right foot contact and one left foot contact) \cite{holden2017pfnn}. Since many stylised motions do not contain a global phase, that is they are not globally cyclic, with different joints moving with different timings and speeds, we choose to use a local phase approach similar to \citet{starke2020local}. In this setup the gating network blends the expert weights based on the phase cycles of individually specified joints.  However, in \citet{starke2020local} the local phases are defined based on the contacts of the hands and feet within the environment, for stylised locomotion we do not always have such contact information, so it is necessary to design a contact-free local phase.

To calculate local phases we fit a sinusoidal curve to some source function, $G(t)$, as in \citet[Section 5.2]{starke2020local}. As this fitting process allows us to extract phase values from many possible functions, the question of local phase extraction is really a question of source function design. We use two different source functions, one when contact information is available and another when it is not.

\vspace{-1mm}
\subsubsection{Contact-Based Local Phases}
To create the contact-based source function we automatically extract approximate foot contact labels by thresholding distance and velocity for the foot bones in each frame. If both the ball centred on the bone with radius $d_{max}$ intersects the floor plane and the velocity magnitude of the bone is less than $v_{max}$ we label the frame with a contact $(1)$ and otherwise with no contact $(0)$. We use default values of $d_{max}=0.01$ and $v_{max}=0.15$ which can be altered for a specific motion if the contact detection is overly or insufficiently sensitive. This creates a step function which is $1$ when a contact is present and $0$ otherwise.

\vspace{-1mm}
\subsubsection{Contact-Free Local Phases}
In the contact-free setting we divide the data into the different gaits shown in Table~\ref{tab:breakdown}. For a given bone, for each style and gait, we then calculate the three principal components of its position in the local coordinate frame. To extract a good local phase we are aiming to capture some cyclic pattern formed by the bone's motion, in particular we do not care about the average position of the bone in the character's coordinate frame, so we centre the data using the mean position for that style and gait. We do not standardise by the variance as the variance in position is what we want the principal components to capture. 

In order to create the contact-free source function we project the position of the bone onto the first principal component. This gives us a one-dimensional function which is large in absolute value when the bone is far away from its mean position (in the direction of the first principal component), in particular this creates an approximately sinusoidal function for motions that are locally cyclic. A final subtlety is that we additionally require that the principal components are directed. To see why, imagine we have two bones moving in a cyclic pattern either in sync  (e.g.\ arms swinging back and forth together) or out of sync by half a cycle  (e.g.\ left arm swings forward as right arm swings back). If the principal components are undirected it may be impossible to tell the difference between these two situations from the source function alone. To direct the principal component, we simply check if it has a positive value along the character's forward facing axis and flip the component if it does not.

\vspace{-1mm}
\subsubsection{Processing and Calculating}
To get the final source functions we additionally postprocess the functions described above. Following \citet{starke2020local}, we normalise within a one second window and smoothe with a Butterworth filter. For the contact-based source function this smoothes out the steps, and for the contact-free function reduces any noise created by small variations in movements.

After applying the fitting process we create a function of the form $a_i \cdot \sin (f_i \cdot t - s_i) + b_i $, where $i$ is the frame index, $t$ the timestamp input to source function $G(t)$, and $\{a_i, f_i, s_i, b_i\}$ parameters optimised by the fitting process. From this we can extract a phase value for each frame, $\phi_i = (f_i \cdot t - s_i) \mod 2\pi$. As we wish to take into account motions with limited bone movement, that is there is no clear phase cycle, we additionally scale the phase by the bone velocity, $||\mathbf{v}_w||$, taken over a one second window, $w$. Combining this with the $2D$ representation used by \citet{starke2020local}, for a source function for bone $b$ we can extract a $2D$ phase feature,
\begin{equation}
    \vspace{-2mm}
    \mathbf{p}_i^{(b)} = ||\mathbf{v}_w|| \; \cdot \; a_i \cdot 
    \begin{pmatrix}
        \sin \phi_i \\
        \cos \phi_i \\
    \end{pmatrix}.
    \label{eq:localphase}
\end{equation}

\begin{figure}
    \centering
    \includegraphics[width=\linewidth]{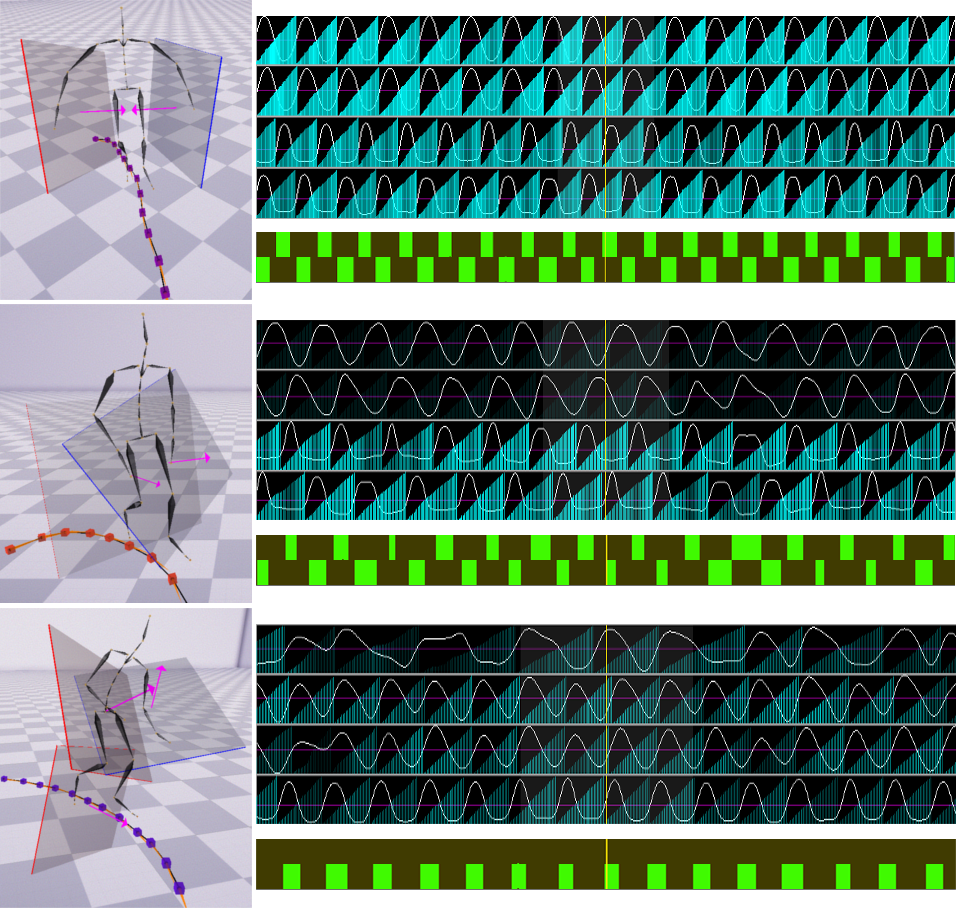}
    \caption{Local Phase Labelling. Top to bottom the styles are: \emph{Pendulum Hands} (local phases are well captured, both contact-free and contact-based); \emph{Hands In Pockets} (low phase magnitude in the hands due to limited movement), and \emph{Left Hopping} (using the contact-free method for the right foot). The yellow line marks the frame shown in the images. The source functions top to bottom are for the right hand, left hand, right foot and left foot.}
    \label{fig:local_phase}
    \vspace{-5mm}
\end{figure}

We calculate local phases in this manner for the end effector bones (hands and feet), using contacts when they are available. The results of this process are visualised in Figure~\ref{fig:local_phase} which shows frames from three different styles. Planes perpendicular to the first principal component are shown for the bones using the contact-free method with the directions shown with purple arrows. Also shown are corresponding source functions (white curves) and extracted phases (blue lines, transparency representing phase magnitude) alongside foot contact detections (green bars). 

\vspace{-2mm}
\subsection{Style Modulation Network}
By using the stylised motion capture clips as conditioning input we can learn a function to parameterise locomotion style. Our style modulation network uses a FiLM generator \cite{perez2018film}, $\bm{\Psi}$, that learns to parameterise style as FiLM parameters which modulate the hidden layers of the motion synthesis network with an affine transformation. We additionally apply layer normalisation \cite{ba2016layer} before the FiLM parameters, creating a form of conditional normalisation \cite{dumoulin2017style}. For a single input at frame $j$, $\mathbf{x}_j$, we can describe our system mathematically as follows. The FiLM generator outputs FiLM parameters for each hidden layer of the motion synthesis network $\bm{\Phi}$,
\begin{equation}
\vspace{-1mm}
    \bm{\Psi}(\mathbf{y}) = \left\{ \bm{\gamma}^{(1)}, \bm{\beta}^{(1)}, \bm{\gamma}^{(2)}, \bm{\beta}^{(2)} \right\},
\label{eq:film_gen}
\vspace{-1mm}
\end{equation}
where $\bm{\gamma}^{(i)}$ and $\bm{\beta}^{(i)}$ are the FiLM parameter vectors for layer $i$. These parameters element-wise scale and shift the hidden units of $\bm{\Phi}$,
\begin{equation}
\vspace{-1mm}
\mathbf{h}^{(i)}_j = ELU \left( \bm{\gamma}^{(i)} \circ \left[ \frac{\bm{F}^{(i)}_j - \bm{\mu}_j}{\bm{\sigma}_j} \right] + \bm{\beta}^{(i)} \right),
\label{eq:modulate}
\vspace{-1mm}
\end{equation}
where $\mathbf{F}^{(i)}_j$ is the output features from layer $i$ of $\bm{\Phi}$ and $\mathbf{h}^{(i)}_j$ is the input to layer $i+1$, for input $\mathbf{x}_j$. $ELU$ is the exponential liner unit \cite{clevert2015elu}, $\circ$ the element-wise product and $\bm{\mu}_j, \bm{\sigma}_j$ the layer normalisation statistics found by calculating the mean and standard deviation over the features for training example $\mathbf{x}_j$. Note that $i \in \{1,2\}$, in particular we only apply the conditional normalisation to the hidden layers. The dashed red box in Figure~\ref{fig:architecture} represents calculating the FiLM parameters as in Equation~\ref{eq:film_gen} and the green boxes between layers of the motion synthesis network capture Equation~\ref{eq:modulate}.

\vspace{-2mm}
\subsection{Combining Animation Synthesis and Style}

We combine animation synthesis and style modulation to predict an output frame $\mathbf{z}_j^s$ from input $\mathbf{x}_j^s$ for style $s$ as:

\begin{equation}
\vspace{-1mm}
\hat{\mathbf{z}}_j^s = \bm{\Phi}\left( \mathbf{x}_j^s; \bm{\Omega}(\mathbf{p}_j^{(1)}, \mathbf{p}_j^{(2)}, \mathbf{p}_j^{(3)}, \mathbf{p}_j^{(4)}), \bm{\Psi}(\mathbf{y^s}) \right),
\label{eq:pred}
\end{equation}
where $\hat{\mathbf{z}}_j^s$ is the prediction, $\bm{\Phi}$ the motion synthesis network whose parameters are experts gated by $\bm{\Omega}$ and whose hidden representations are modulated with $\bm{\Psi}$. $\mathbf{p}_j^{(i)}$ are the end effector phase features for frame $j$ (Equation~\ref{eq:localphase}) and $\mathbf{y^s}$ a stylised motion clip for style $s$.

The final parameters for $\bm{\Phi}$ form three feed-forward layers with both hidden layers being of dimension $512$. The gating network is also three feed-forward layers with both hidden layers of size $32$ and $8$ output units, for the $8$ experts we blend. Our FiLM generator, $\bm{\Psi}$, consists of two 1-dimensional convolutional layers containing 256 filters of size 25 that convolve over the temporal dimension each followed by max pooling, downsampling by a factor of 2 each time. We then apply two fully connected layers with 2048 hidden units and output units. The output units correspond to the four 512-dimensional vectors from Equation \ref{eq:film_gen}.

To train the model we select 95 styles from the \ostyle\ dataset, keeping the remaining styles for testing fine-tuning performance on new styles (Section~\ref{sec:analysis_extensions}). The system is trained end to end using mean squared error across all predicted features in $\hat{\mathbf{z}}_j$ and regularised with a dropout rate of $0.3$ everywhere \cite{srivastava2014dropout}. We also find a bone length loss useful to help reduce bone stretching artifacts. This creates our objective function $\mathcal{L} = \mathcal{L}^{mse} + \mathcal{L}^{bll}$ where:
\vspace{-1mm}
\begin{dgroup}
\begin{dmath}
\mathcal{L}^{mse} =  \sum_{s=1}^{S} \frac{1}{N_s} \sum_{j=1}^{N_s}  (\mathbf{z}_j^s - \hat{\mathbf{z}}_j^s)^2 
\label{eq:mse}
\end{dmath}
\begin{dmath}
l^{bll}(\mathbf{z}, \hat{\mathbf{z}}) = \frac{1}{B} \sum_{b=1}^B \left| ||z_{b_p} - z_b ||_2 - ||\hat{z}_{b_p} - \hat{z}_b ||_2 \right|
\label{eq:individual_bll}
\end{dmath}
\begin{dmath}
\mathcal{L}^{bll} = \sum_{s=1}^{S} \frac{1}{N_s} \sum_{j=1}^{N_s} l^{bll}(\mathbf{z}_j^s, \hat{\mathbf{z}}_j^s) 
\label{eq:bll}
\end{dmath}
\end{dgroup}
\vspace{-1mm}
where $\mathbf{z}_j^s$ is the ground truth output for frame $j$ in style $s$,  $\hat{\mathbf{z}}_j^s$ is the corresponding network prediction, $S$ is the total number of training styles, $N_s$ is the number of training frames available for style $s$, $B$ is the number of bones in the skeleton, $z_b$ is the $3D$ position vector for the child joint of a bone for ground truth $\mathbf{z}$, $z_{b_p}$ is the parent joint position vector, and $\hat{z}_{b}$ \& $\hat{z}_{b_p}$ are the corresponding vectors in the predicted $\mathbf{\hat{z}}$. We use minibatch training and, to ensure balanced training, each minibatch contains only one style and we cycle through all the styles in a fixed order, so every set of $S$ minibatches contains all of the training styles. As styles may have different numbers of training frames we define one epoch as one pass through the shortest dataset and train for 90 epochs using Adam \cite{kingma2014adam} with a learning rate of $1 \times 10^{-4}$ on a single Nvidia GeForce GTX 1080 Ti.

\vspace{-2mm}
\subsection{Creating the Real-Time Demo}
\label{sec:engineering}

We create a real-time demo which allows a user to control a character with keyboard input. We follow the method of \citet{zhang2018mann} to create a trajectory from user input which is blended with the network's predicted future trajectory. In order to reduce foot sliding artifacts we use the predicted foot contact label for the output frame $\mathbf{\hat{z}}$ to post-process our animation with foot contact inverse kinematics. We skin our character with the Mixamo~\cite{mixamo} Y bot skin. We additionally find that when interpolating between styles the predicted rotations can sometimes be inaccurate causing the character skin to become deformed. This is likely because linear interpolation with our rotation representation does not always correspond to the correct rotation interpolation and learning this non-linear interpolation with linear FiLM interpolation is difficult. To account for this we apply inverse kinematics along each arm to map the bones to the predicted positions which ensures the rotations reasonably match these positions.

To increase speed and reduce storage costs at runtime we can pre-calculate and save the FiLM parameters for each style offline, this allows us to save a single vector (see Section~\ref{sec:costs}) either as an average of all FiLM parameters or from a single style clip. To interpolate between styles we allow the user to select three arbitrary training styles and blend the FiLM parameters according to the barycentric coordinates at a triangle location selected by the user. These interpolations can be manipulated online.

\section{Evaluation}
\label{sec:evaluation}

Equipped with our new dataset, the variety of styles available allows us to: highlight failure cases with other style modelling approaches that we are able to resolve; analyse our methodology and, in Section~\ref{sec:discussion}, motivate future work in style modelling.

Our main comparisons from the literature are \emph{PFNN One-hot} \citep{holden2017pfnn} and \emph{PFNN Resad} \citep{mason2018style}, which use the PFNN for animation synthesis and modulate style using a one-hot vector and residual adapters \citep{rebuffi2017residual} respectively. We also compare animation synthesis backbones, utilising FiLM style modulation with the PFNN (\emph{PFNN FiLM}), MANN (\emph{MANN FiLM}) \cite{zhang2018mann} and LPN (\emph{LPN FiLM}). The latter is our system, outlined in Section~\ref{sec:method} . Finally, we combine the different style modulation approaches with our LPN (\emph{LPN One-hot} \& \emph{LPN Resad}).

\vspace{-2mm}
\subsection{Reducing Failure Cases}

Initially we evaluate the different methods qualitatively, highlighting different ways in which methods can struggle. These results are additionally shown, and best viewed, in the supplementary video.
\begin{figure}[h]
    \centering
    \setlength{\abovecaptionskip}{5pt plus 2pt minus 2pt}
    \includegraphics[trim={0 1cm 0 0.5cm}, clip, width=\linewidth]{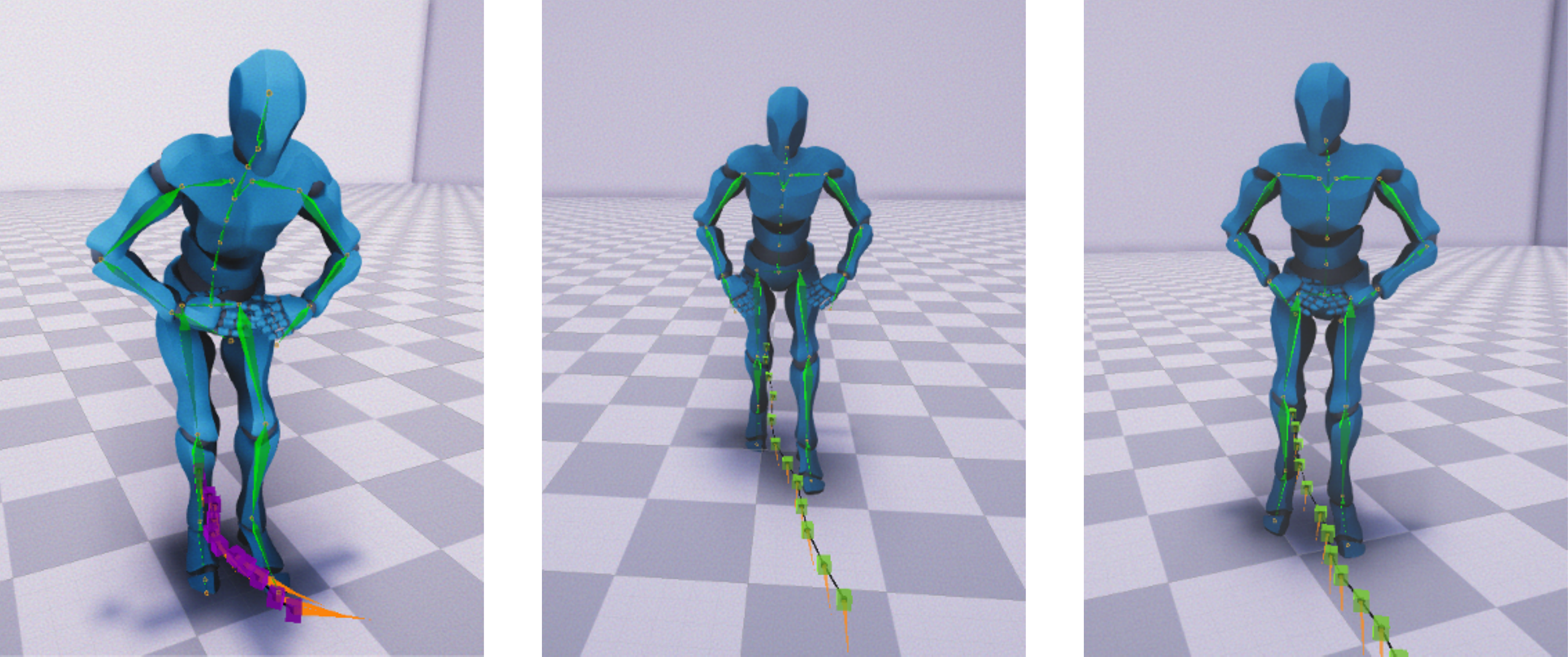}
    \caption{No Global Phase. Style: \emph{Pendulum Hands}. Left to Right: Data, PFNN FiLM, LPN FiLM. For the PFNN the hands do not swing together correctly.}
    \label{fig:no_phase}
    \vspace{-4mm}
\end{figure}
\subsubsection{Motions Without a Global Phase}
As in \citep{starke2020local} we find that the introduction of a local phase is extremely beneficial for motions which have no clear single global phase variable. For many styles, the models which use the PFNN for animation synthesis struggle to capture the movement of all body parts correctly. Since the PFNN's global phase is extracted from foot contacts this often manifests as the hands and arms of the character becoming `stuck' or performing some average pose as shown in Figure~\ref{fig:no_phase}. The challenges with modelling styles without coherent global phases emphasise the importance of using a more flexible animation synthesis architecture.

\begin{figure}[h]
    \centering
    \setlength{\abovecaptionskip}{5pt plus 2pt minus 2pt}
    \includegraphics[trim={0 1.5cm 0 0}, clip, width=\linewidth]{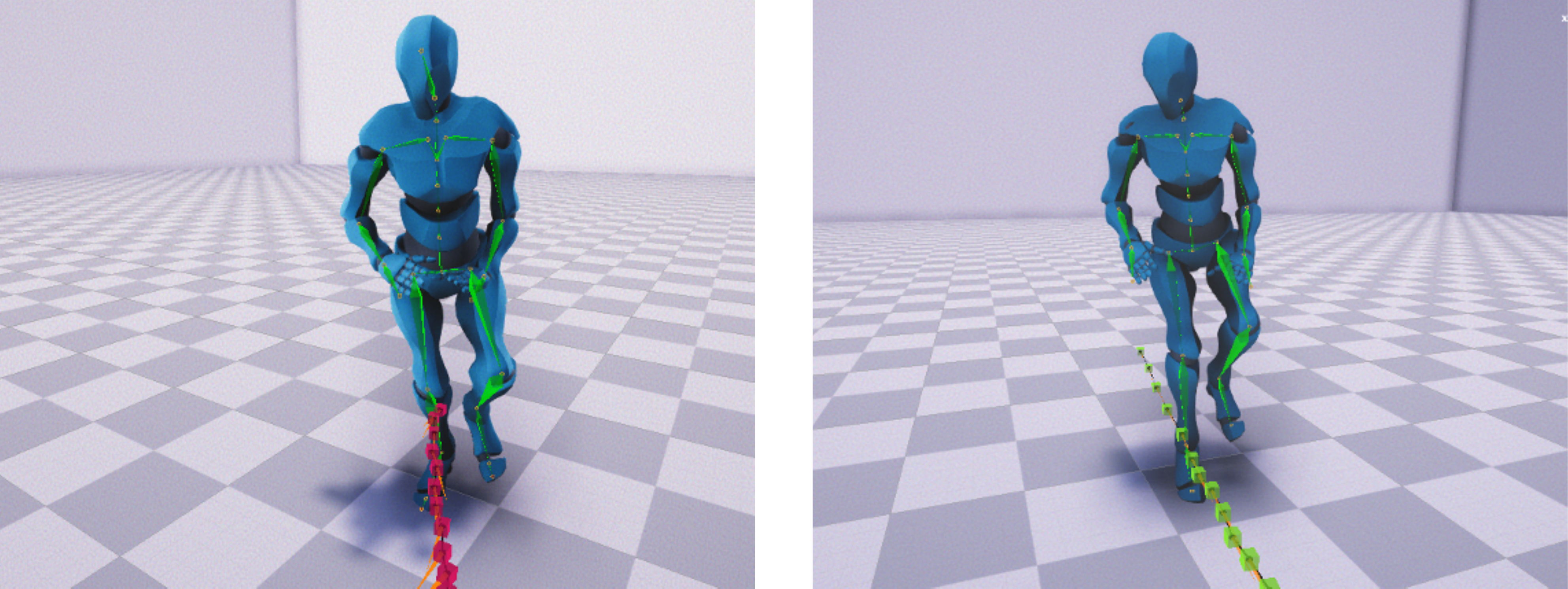}
    \caption{Contact-Free Modelling. Style: \emph{Left Hop}. Left: Data. Right: LPN FiLM.}
    \label{fig:no_contact}
    \vspace{-4mm}
\end{figure}

\paragraph{Contact-Free Modelling}
In Section~\ref{sec:anim_syn} we formulated a method to create contact-free local phases, this easily allows for styles that have varying contacts with the environment. Figure~\ref{fig:no_contact} shows a style where only a single foot makes contact with the ground.

\begin{figure}[h]
    \centering
    \setlength{\abovecaptionskip}{5pt plus 2pt minus 2pt}
    \includegraphics[trim={0 2cm 0 1.5cm}, clip, width=\linewidth]{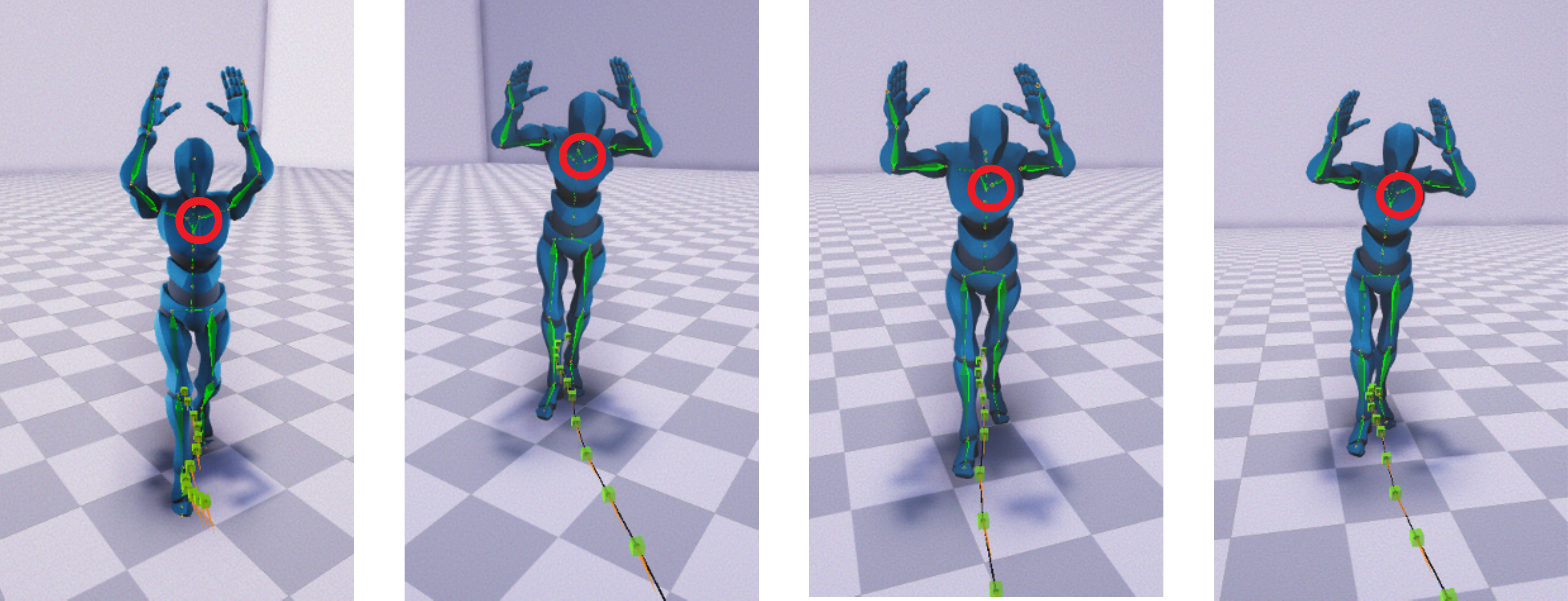}
    \caption{Capacity Challenges. Style: \emph{Arms Above Head}. Left to Right: Data, LPN One-hot, LPN Resad, LPN Film. The circled clavicle bones are poorly modelled with lower capacity.}
    \label{fig:capacity}
    \vspace{-4mm}
\end{figure}

\subsubsection{Dealing With Large Numbers of Styles}
\label{sec:capacity}
When modelling all styles from the \ostyle\ dataset we find that the lower capacity of one-hot and residual adapter modulation manifests in animation artifacts (Figure~\ref{fig:capacity}).
\looseness-1 In particular, as implemented in the original works, both the one-hot and residual adapter methods can be seen as adding a style-dependent shift to a \emph{single} hidden layer of the synthesis network, whereas FiLM both scales and shifts in \emph{all} hidden layers. We note also that in the one-hot case this problem can additionally cause styles to be modelled incorrectly (see supplementary video).

\begin{figure}[h]
    \centering
    \setlength{\abovecaptionskip}{5pt plus 2pt minus 2pt}
    \includegraphics[trim={0 3.5cm 0 0.5cm}, clip, width=\linewidth]{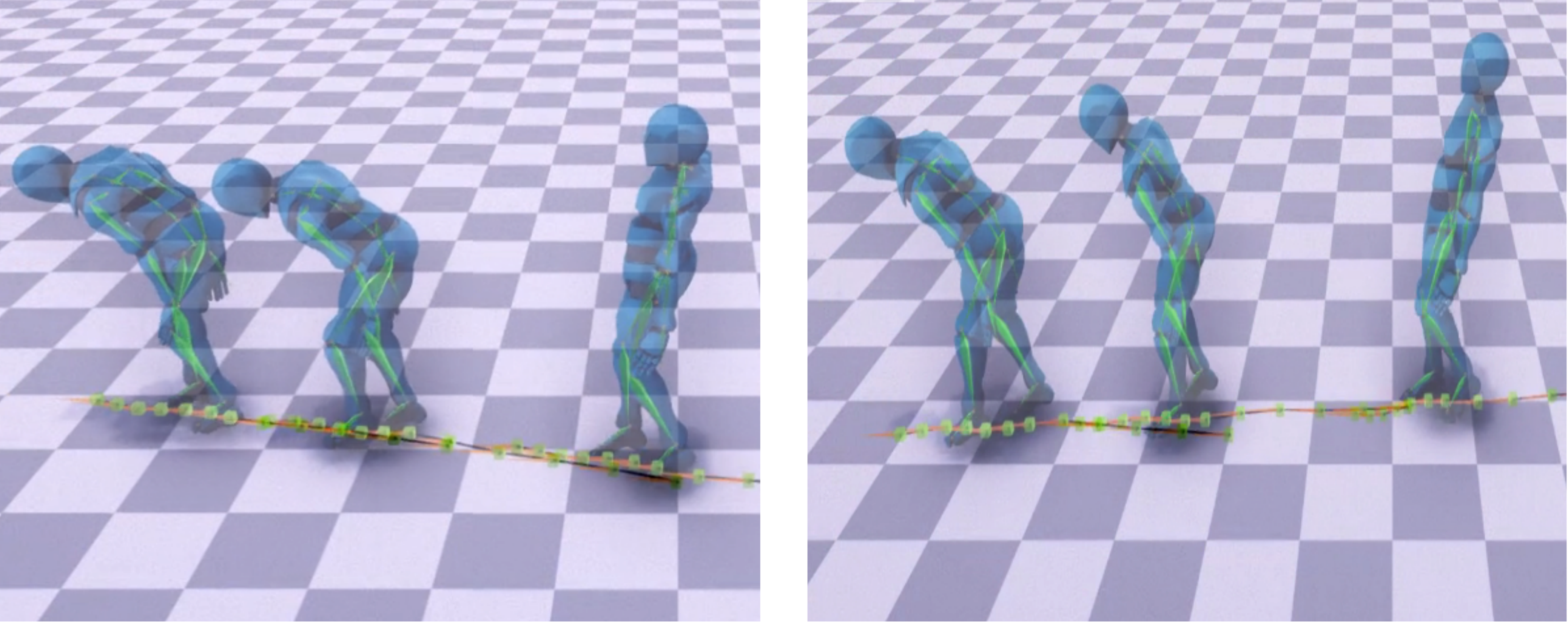}
    \caption{Interpolation. Style: \emph{Bent Forward} to \emph{Lean Back}. Left: LPN One-hot. Right: LPN FiLM. The one-hot method does not smoothly blend between styles.}
    \label{fig:interpolation}
    \vspace{-4mm}
\end{figure}

\paragraph{Interpolation Difficulties}
We additionally find that when interpolating between styles the one-hot approach fails to learn a smoothly interpolatable representation. Rather, as we move between two styles, this method tends to maintain the first style until interpolation is complete. On the other hand, FiLM learns a continuous style space to give a smooth interpolation as shown in Figure~\ref{fig:interpolation}. This preferable interpolation behaviour alongside the artifacts highlighted above evidence our preference for FiLM style modulation.

\begin{figure}[h]
    \centering
    \setlength{\abovecaptionskip}{5pt plus 2pt minus 2pt}
    \includegraphics[trim={0 2cm 0 1.5cm}, clip, width=\linewidth]{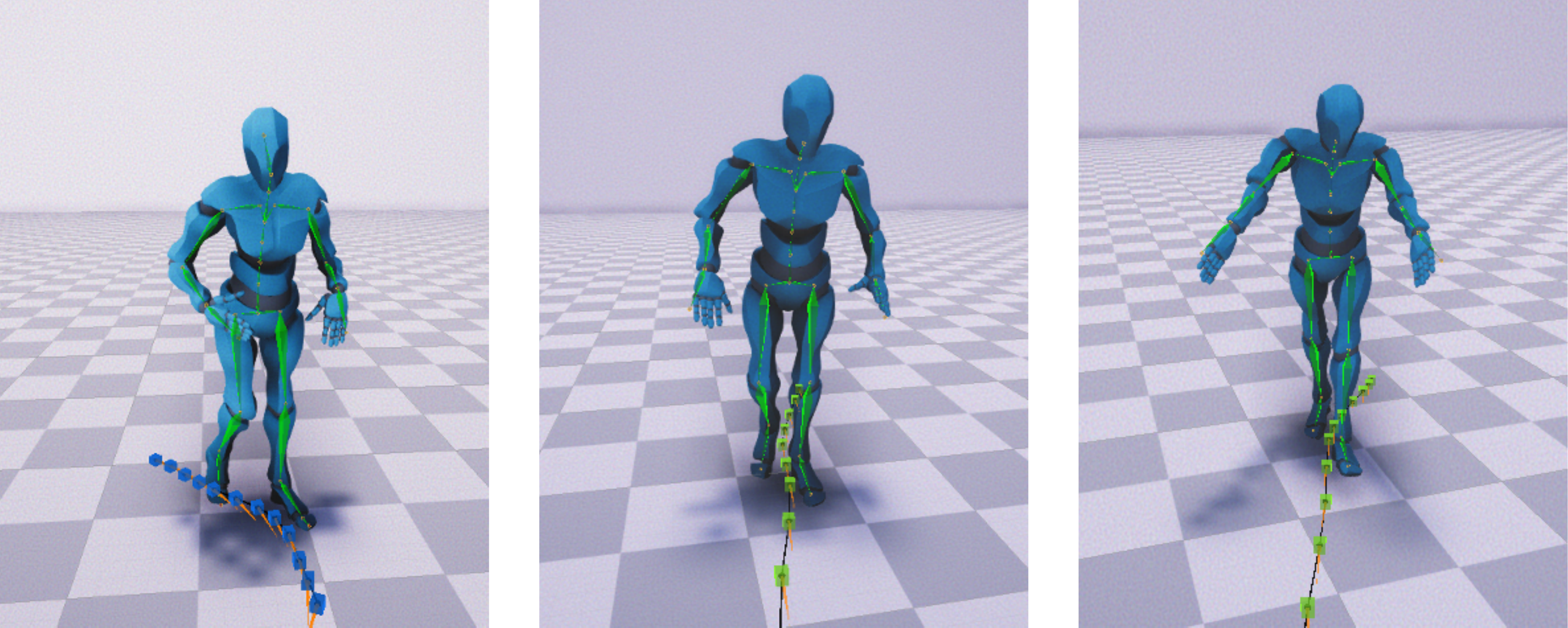}
    \caption{No Phase Features. Style: \emph{Swimming}. Left to Right: Data, MANN FiLM, LPN FiLM. Note the arms get stuck in front of the character when using the MANN.}
    \label{fig:no_phase_feature}
    \vspace{-4mm}
\end{figure}

\subsubsection{No Explicit Phase Features}
Finally, we consider the MANN which uses end effector velocities as gating input. We find this method to be competitive, however, like the PFNN, it can struggle when limbs have distinct phase cycles (Figure~\ref{fig:no_phase_feature}). We believe that the LPN network's use of explicit phase features for gating allow it to separate of such styles more easily. This also further validates the utility of designing source functions for local phases.

\vspace{-2mm}
\subsection{Numerical Comparisons}
\label{sec:costs}
We now turn to a brief quantitative analysis of our models.

\subsubsection{Computational Costs}

\begin{table}[]
\caption{Storage and Runtime Comparisons. \emph{A.S.N.}, \emph{S.M.N.} \& \emph{P.S.R.} show parameter counts for the animation synthesis network, style modulation network and per style at runtime. Runtime (ms) shows the mean and standard deviation of the average time to calculate the next frame over five runs.}
\vspace{-2mm}
\begin{tabular}{@{}lllll@{}}
\toprule
Model        & A.S.N.      & S.M.N.      & P.S.R.  & Runtime (ms)  \\ \midrule
PFNN One-hot & 2,436,380    & 194,560     & 2048          & $0.78\pm0.01$ \\
LPN One-hot  & 4,935,928   & 389,120     & 4096          & $4.11\pm0.02$ \\
PFNN Resad   & 2,436,380   & 2,978,440   & 31,352        & $0.92\pm0.01$ \\
LPN Resad    & 4,935,928   & 24,952,320  & 262,656       & $4.78\pm0.06$ \\
PFNN FiLM    & 2,436,380   & 35,668,530  & 2048          & $1.37\pm0.02$ \\
MANN FiLM    & 4,870,392   & 35,668,530  & 2048          & $4.48\pm0.04$ \\
LPN FiLM     & 4,935,928   & 35,668,530  & 2048          & $4.53\pm0.04$ \\ \bottomrule
\end{tabular}
\label{tab:costs}
\vspace{-4mm}
\end{table}

Initially we consider the memory footprint and runtime of each of the models, see Table~\ref{tab:costs}. Although all models remain comfortably able to perform at $60$fps, the improved modelling of the LPN (and MANN) comes with an additional runtime cost. Whilst these differences can be partly attributed to differing parameter counts they are also in part due to an optimisation of the PFNN that scales poorly for gated expert systems. So in situations with small numbers of globally cyclic movements, using the PFNN as the animation synthesis network may be preferable.

Arguably the most important column of Table~\ref{tab:costs} is P.S.R. -  the parameters per style at runtime. This highlights that, in practice, FiLM does not require its high number of style modulation parameters (S.M.N. column) since the output can be pre-computed (Section~\ref{sec:engineering}). This is in contrast to residual adapters, for which all parameters must be saved as their input will change during inference. Furthermore, that LPN One-hot uses twice as many parameters per style as LPN FiLM demonstrates the importance of the design choices for how to apply the style representation to modulate animation synthesis. This low cost for style representation additionally allows us to cheaply save the pre-computed FiLM parameters and then fine-tune the FilM generator to model new styles (see Section~\ref{sec:analysis_extensions}).

\subsubsection{Test Error}

\begin{figure}[h]
    \centering
    \includegraphics[width=0.6\linewidth]{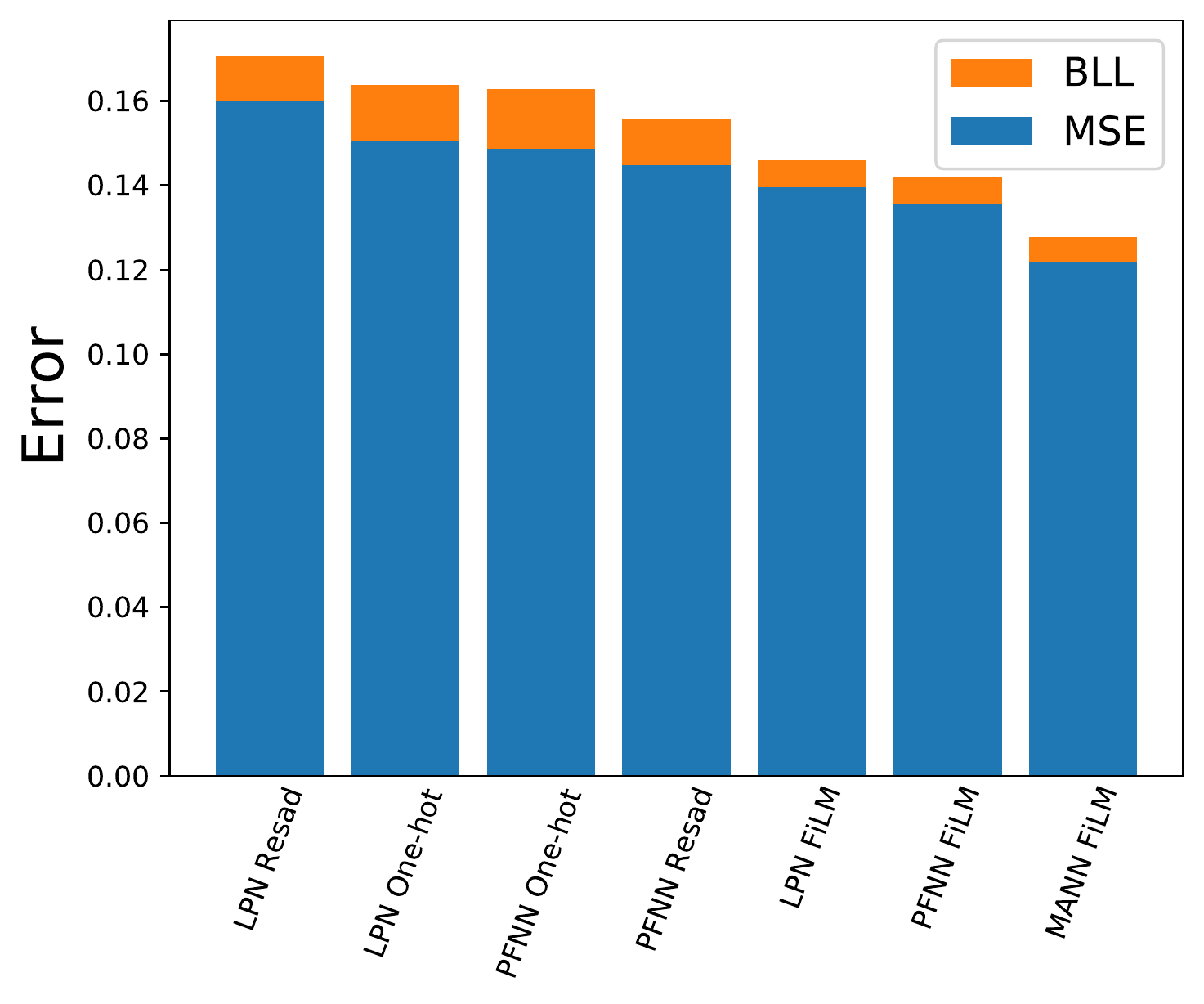}
    \vspace{-4mm}
    \caption{Test Error. The error is divided into the mean squared error (MSE) and the bone length loss (BLL).}
    \label{fig:error}
    \vspace{-4mm}
\end{figure}

Figure~\ref{fig:error} shows test errors using seen styles and unseen frames of motion. Our overwhelming finding is, like \citet{Martinez_2017_CVPR}, that mean squared error does not correlate strongly with perceived qualitative performance (performing a `stuck' average motion, can give very low average error). However, whilst this does mean that one must take care not to over-interpret the results, it does not mean that there is no insight to be gained.

What we do see in Figure~\ref{fig:error} is that all the models using FiLM have the lowest test errors which we again attribute to the increased capacity of FiLM for changing the hidden features, as discussed above. Furthermore, the difference between LPN Resad and PFNN Resad suggests an importance for the CP decomposition \cite{mason2018style} which adds a phase dependence to the residual adapter (which we did not replicate in the LPN due to the lack of a global phase). These observations further emphasise that, in this situation, the inductive biases underlying the choices for how to use the parameters available are more important than the raw network capacity.

\vspace{-2mm}
\subsection{Analysis and Extensions}
\label{sec:analysis_extensions}

Finally, we analyse our model within a wider theoretical context and evaluate a method for extending our system to new styles.

\subsubsection{Theoretical Analysis.}
Both FiLM and residual adaptation can be seen as a special case of the Multi-Task Dynamical Systems (MTDS) framework which models style by changing the parameters of neural network layers \cite{bird2019multitask}.  
To see this, we write a feed-forward layer as $\bm{h}^{(i+1)} = \sigma ( \bm{W} \bm{h}^{(i)} + \bm{b} )$,
with $\sigma$ a non-linearity, $\bm{h}^{(i)}$ the hidden units for layer $i$, and $\bm{W}, \bm{b}$ the layer weights and biases. If we then represent the element-wise FiLM scaling for style $s$ using diagonal matrix $\bm{W}^s$, with the individual scaling factors on the diagonal, and keep the FiLM shift vector $\bm{\beta}^s$, we can rewrite FiLM as:

\begin{equation}
    \bm{h}^{(i+1)} =  \sigma \left( \bm{W}^s (\bm{W} \bm{h}^{(i)} + \bm{b}) + \bm{\beta}^s \right) =  \sigma \left( \bm{W}' \bm{h}^{(i)} + \bm{b}' \right),
    \label{eq:bird}
\end{equation}

\noindent where $\bm{W}' = \bm{W}^s \bm{W}$ and $\bm{b}' = \bm{W}^s \bm{b} + \bm{\beta}^s$ are the new parameters - a specific parameter manipulation, as done more generally in MTDS.

However, capturing styles with weight matrices is expensive. Sharing parameters across all styles allows us to be more compact with our style representation, as style agnostic variation can be modelled with one shared set of parameters. Compared to other methods \cite{bird2019multitask, mason2018style}, our style representation is significantly smaller.

\subsubsection{Modelling New Styles}

Despite the size of the \ostyle\ dataset, training does not give a dense enough sample of the style space to achieve robust generalisation to unseen styles \cite{ji2021test}. A natural question then, is how we could model new styles with our system.

We can view LPN parameters as a style agnostic motion representation to which we simply need add the correct representation for the new style. Assuming we have training data for a new style we can simply fine-tune the FiLM generator (red box in Figure~\ref{fig:architecture}) while freezing the parameters of the LPN to extract such a representation. Note that for the one-hot approach, learning a new style would require retraining all network parameters, and for residual adapters would require storing another large set of parameters (Table~\ref{tab:costs}).

\begin{figure}
    \centering
    \setlength{\abovecaptionskip}{5pt plus 2pt minus 2pt}
    \begin{subfigure}{.5\linewidth}
        \centering
        \includegraphics[width=\linewidth]{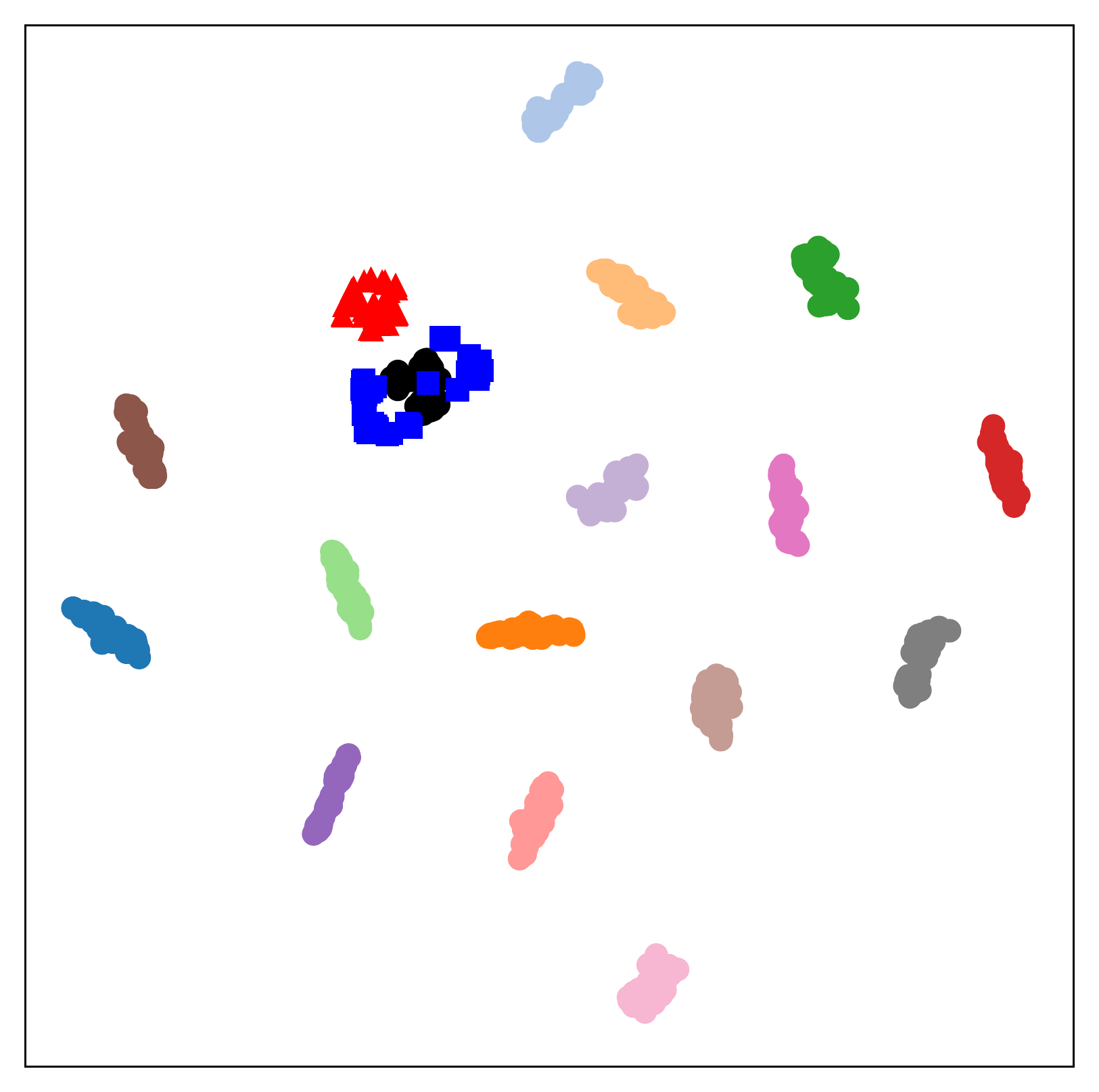}
        \caption{Before Fine-Tuning}
        \label{fig:tsne:generalisation}
    \end{subfigure}%
    \begin{subfigure}{.5\linewidth}
        \centering
        \includegraphics[width=\linewidth]{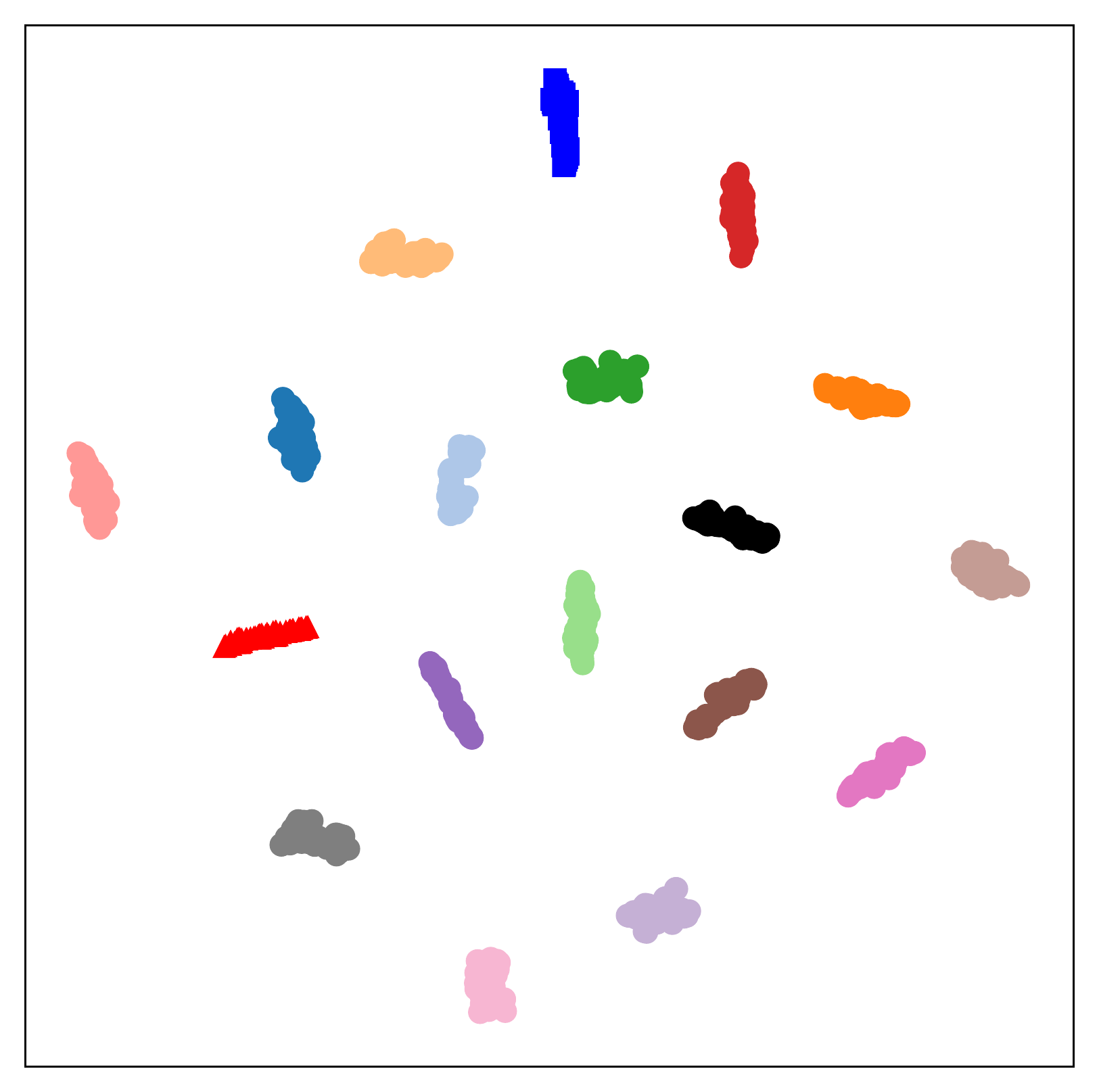}
        \caption{After Fine-Tuning}
        \label{fig:tsne:finetuning}
    \end{subfigure}
    \caption{t-SNE. FiLM parameters calculated from 50 clips of each style are clustered with t-SNE \cite{maaten2008tsne}. Shown are 15 styles from the training set (pastel colours) and 3 unseen styles (black circles, blue squares, red triangles). After fine-tuning the unseen styles show much clearer separation.}
    \label{fig:tsne}
    \vspace{-3mm}
\end{figure}

Figure~\ref{fig:tsne} gives some intuition for why this approach makes sense. In Figure~\ref{fig:tsne:generalisation} FiLM parameters for different styles before performing fine-tuning are visualised; we see clear separation for the styles the network was trained on, but for unseen styles the separation is worse, suggesting the FiLM generator has not extracted specific style representations. After fine-tuning (Figure~\ref{fig:tsne:finetuning}), we see that unseen styles are well separated at their own location in the style space. Qualitative results are shown in the supplementary video.

\section{Discussion}
\label{sec:discussion}

From a single stylised motion clip we can now generate arbitrary length user controlled motions in a variety of styles (Figure~\ref{fig:teaser}). In this section we highlight arising areas useful for further discussion.

\paragraph{What is Style Anyway?} By using a shared set of LPN parameters our model implicitly assumes motion content to be `the things that are the same across styles in the dataset' and motion style to be `all the remaining things that differ', that is we make a content invariance assumption \cite{vk2021content}. However, this is not the only possible definition of style which is a subjective and creative concept. We hope that this work may be useful for future creative systems.

\paragraph{Source Function Design} In Section~\ref{sec:method} we saw that the extraction of local phases is really a question of source function design. We have provided one possible source function that works for our use case, but we believe there is a large scope for designing different source functions for different tasks. For example, our source function does not handle stochasticity well, so designing contact-free techniques for new situations is be a useful direction for further exploration.

\paragraph{Physics Understanding} Finally, whilst our model interpolates smoothly, it often takes a direct path between motion styles which can lead to physically implausible poses (clipping). The constrained  generality of human motion almost certainly requires some understanding of physics and further investigating systems that learn world models \cite{ha2018world} to encourage physics understanding \cite{fussell2021supertrack}, or directly adding skeletal awareness to neural networks \cite{aberman2020skeleton}, may allow us to generalise future models over the entire motion space.

\section{Conclusion}

This work has explored the style modelling task for humanoid animation. By creating and releasing the new \ostyle\ dataset, we were able to pose challenges to existing systems and motivate solutions in the form of (contact-free) local phases and higher capacity feature-wise linear transformations. Our compact but powerful style representation showed meaningful linear interpolations in the learned style space and, by improving animation synthesis, we were able to model a wider variety of motions.

\section*{Acknowledgements}
We thank Alex Bird and Chris Williams for useful input on the MTDS framework, Cian Eastwood for providing feedback on an early draft, and numerous colleagues for useful discussions. IM is supported by the Engineering and Physical Sciences Research Council (EPSRC).

\bibliographystyle{ACM-Reference-Format}
\bibliography{bibliography}

\end{document}